\documentclass[preprint, superscriptaddress, showpacs,preprintnumbers,amsmath,amssymb,pra]{revtex4}
\usepackage{graphicx}

\begin{document}

\thispagestyle{empty}

\title{Optical properties of dielectric plates coated with
gapped graphene}

\author{
G.~L.~Klimchitskaya}
\affiliation{Central Astronomical Observatory at Pulkovo of the
Russian Academy of Sciences,
Saint Petersburg,
196140, Russia}
\affiliation{Institute of Physics, Nanotechnology and
Telecommunications, Peter the Great Saint Petersburg
Polytechnic University, St.Petersburg, 195251, Russia}

\author{
V.~M.~Mostepanenko}
\affiliation{Central Astronomical Observatory at Pulkovo of the Russian Academy of Sciences,
Saint Petersburg,
196140, Russia}
\affiliation{Institute of Physics, Nanotechnology and
Telecommunications, Peter the Great Saint Petersburg
Polytechnic University, St.Petersburg, 195251, Russia}
\affiliation{Kazan Federal University, Kazan, 420008, Russia}

\begin{abstract}
The optical properties of dielectric plates coated with gapped
graphene are investigated on the basis of first principles of
quantum electrodynamics. The reflection coefficients and
reflectivities of graphene-coated plates are expressed in terms
of the polarization tensor of gapped graphene and the dielectric
permittivity of plate material. Simple approximate expressions
for the required combinations of components of the polarization
tensor applicable in the wide frequency region, where the
presence of a gap influences the optical properties, are found.
Numerical computations of the reflectivities of graphene-coated
SiO${}_2$ plates are performed for different values of the
mass-gap parameter at different temperatures. It is shown that
with an increasing gap width the reflectivity of a
graphene-coated plate at the normal incidence decreases by up
to a factor of 8 depending on the values of frequency and
mass-gap parameter. The angle dependences of reflectivities
for both polarizations of the incident electromagnetic waves
have been computed for Si and SiO${}_2$ plates coated with
gapped graphene. We demonstrate that the TM reflectivity has
a minimum value at some angle of incidence depending on the
mass-gap parameter, frequency and temperature, whereas the TE
reflectivity depends on the angle of incidence monotonously.
However, for the graphene coatings with a nonzero mass-gap
parameter the reflected light cannot be fully polarized.
Possible applications of the obtained results are discussed.
\end{abstract}

\maketitle

\section{Introduction}

Graphene is one of the family of two-dimensional materials
possessing unusual optical, electrical and mechanical
properties of much interest for both fundamental physics
and applications in nanotechnology \cite{1}. This is
because at low energies the electronic excitations in
graphene are described not by the Schr\"{o}dinger equation
but by the relativistic Dirac equation, where the speed of
light $c$ is replaced with the Fermi velocity
$v_F\approx c$/300. It has been demonstrated \cite{2,3} that in
the region of visible light the transparency and electrical
conductivity of graphene are defined via the fundamental
physical constants $\alpha=e^{2}/(\hbar c)$ and
$e^{2}/(4\hbar)$, respectively. This was interpreted
theoretically using the transport theory, current-current
correlation functions, and the Kubo formula
(see Refs.~\cite{4,5,6} for a review).

The electrical conductivity of graphene calculated within
the cited approaches has been used to find the reflection
coefficients and reflectivities of graphene with zero
mass-gap parameter $\Delta=2mc^{2}$, where $m$ is the mass
of electronic excitations \cite{7,8}. In so doing, the case
of transverse magnetic (TM, i.e., $p$-polarized)
electromagnetic waves was considered. Here, it is pertinent
to note that for a pure (pristine) graphene $\Delta=m=0$,
but under the influence of defects of the structure,
electron-electron interaction, and for graphene deposited
on a substrate electronic excitations acquire some nonzero
mass \cite{9,10,11,12,12a}. Usually the values of the
mass-gap parameter are estimated as $\Delta\leqslant 0.1$ or
0.2 eV.

On the basis of first principles of quantum electrodynamics,
one can express both the reflectivity and conductivity of
graphene via the polarization tensor in (2+1)-dimensional
space-time. 
For graphene this tensor was found at zero \cite{13} and
nonzero \cite{14} temperature for the values of frequency
restricted to only pure imaginary Matsubara frequencies.
Note that in the framework of (2+1)-dimensional quantum 
electrodynamics with no relation to graphene the polarization 
tensor was studied long ago, e.g., in Refs. \cite{14a,14b}.
The  results of Refs. \cite{13,14}  were applied to investigate the Casimir
force in graphene systems in the framework of the Lifshitz
theory \cite{15,16,17,18,19,22} (previous calculations
on this subject have been made using the density-density
correlation functions, model dielectric permittivities
of graphene and
other methods \cite{23,24,25,26,27,28}).

It is pertinent to note that the Lifshitz theory of the
van der Waals and Casimir forces uses the reflection
coefficients of interacting surfaces at pure imaginary
Matsubara frequencies \cite{28b}. For calculation
of the optical and conductivity properties of graphene,
an analytic continuation of the polarization tensor to real
frequency axis is required. Such a continuation was found
in Ref. \cite{29} and used to study the reflectivity
properties of pure graphene at any temperature over the
wide frequency region on the basis of first principles of
quantum electrodynamics. In Ref. \cite{30} the same
formalism was applied to investigate the electrical
conductivity of pure graphene. Finally, in Ref. \cite{31}
an explicit representation for the polarization tensor of
gapped graphene along the real frequency axis has been
obtained and used for investigation of its reflectivity
properties \cite{31} and electrical conductivity \cite{31a}.
It was shown that a nonzero mass of electronic
excitations has a profound effect on the reflectivity of
graphene at both zero and nonzero temperature \cite{31}.
Specifically, at zero temperature the reflectivities of
gapped graphene go to zero with vanishing temperature,
whereas for a gapless graphene in this limiting case the
reflectivities go to a nonzero constant depending only on
$\alpha$ and on the angle of incidence. At nonzero
temperature, in contrast to the gapless case, the
reflectivities of gapped graphene drop to zero in the
vicinity of some fixed frequency \cite{31}.

Dielectric plates coated with graphene are used in many
technological applications, such as the optical detectors
\cite{32}, solar cells \cite{33}, transparent electrods
\cite{34}, corrosium protection \cite{35}, optical
biosensors \cite{36}, optoelectronic switches \cite{37},
and many others. The reflection coefficients for material
plates coated with graphene were obtained in Ref. \cite{38}
in the random phase approximation. The exact expressions
for both the TM and TE (i.e., for $s$-polarized
electromagnetic waves) reflection coefficients of
graphene-coated plates, where the plate material is
described by the frequency-dependent dielectric permittivity
and graphene by the polarization tensor, are derived in
Ref. \cite{19}. In Ref.~\cite{39} these reflection
coefficients have been used to investigate the reflectivity
properties of material plates coated with the pure (gapless)
graphene. It was shown that at frequencies much smaller than
the thermal frequency the coating with gapless graphene
results in a pronounced (up to an order of magnitude)
increase in the reflectivity of dielectric plates. At high
frequencies (for metallic plates at all frequencies) the
influence of graphene coating on the reflectivity
properties was shown to be rather moderate \cite{39}.

In this paper, we investigate the optical properties of
dielectric plates coated with gapped graphene on the
basis of first principles of quantum electrodynamics.
For this purpose, graphene coating is described by the
polarization tensor with a nonzero mass-gap parameter in
the form derived in Ref.~\cite{31}, whereas the plate
material is described by its dielectric permittivity.
The analytic expressions for both the TM and TE reflection
coefficients and reflectivities of dielectric plates coated
with gapped graphene are presented at frequencies smaller
than the thermal frequency, where the effect of graphene
coating plays an important role.

Numerical computations are performed for SiO${}_2$ plates
which are of frequent use in experiments as substrates for
graphene coating \cite{40,41,42,43}. Specifically, we
calculate the reflectivities at normal incidence of
SiO${}_2$ plates coated with graphene sheets with different
values of the mass-gap parameter at different temperatures.
It is shown that with an increasing gap width the
reflectivity of a graphene-coated substrate decreases. This
decrease may be in several times, as compared to the case of
a coating with gapless graphene, depending on the values of
frequency and mass-gap parameter. According to our results,
an increase of the gap parameter plays qualitatively the
same role as a decrease of temperature for dielectric plates
coated with a gapless graphene. We have also computed the
angle dependences for both the TM and TE reflectivities of
graphene-coated plates. It is shown that the TM
reflectivity has a minimum value at some angle of incidence,
whereas the TE reflectivity depends on the angle of
incidence monotonically. In so doing, for coatings with
gapped graphene the full polarization of reflected light
turns out to be impossible. The angle dependences are found
not only for graphene-coated SiO${}_2$ plates, but for the plates made
of high-resistivity Si , which are also used as substrates
for graphene coating \cite{44}.

The paper is organized as follows. In Sec. II, we present
analytic expressions for the reflection coefficients of
dielectric plates coated with gapped graphene in different
frequency regions. Section III is devoted to the
investigation of influence of a nonzero gap on the
reflectivity of graphene-coated plates at normal incidence.
In Sec. IV the influence of gapped graphene on the angle
dependence of reflectivities is considered. Section V
contains our conclusions and discussion.

\section{Reflection coefficients in different frequency
regions}

We consider thick dielectric plate (semispace) coated
with the sheet of gapped graphene. The plate material is
described by the frequency-dependent dielectric
permittivity $\epsilon(\omega)$ and graphene by the
polarization tensor $\Pi_{kn}(\omega,\theta_i)$, where
$\theta_i$ is the angle of incidence and $k,n$=0,1,2,
taking into account the nonzero gap parameter. The
reflection coefficients of a graphene-coated plate for two
independent polarizations of the electromagnetic field
defined at real frequencies are given by \cite{39}
\begin{eqnarray}
&&
R_{\rm TM}^{(g,p)}(\omega,\theta_i)=\frac{\varepsilon(\omega)\cos\theta_i-
\sqrt{\varepsilon(\omega)-\sin^2\theta_i}\left[1-
\tilde{\Pi}_{00}(\omega,\theta_i)\right]}{\varepsilon(\omega)\cos\theta_i+
\sqrt{\varepsilon(\omega)-\sin^2\theta_i}\left[1+
\tilde{\Pi}_{00}(\omega,\theta_i)\right]},
\nonumber \\
&&\label{eq1}\\[-3mm]
&&
R_{\rm TE}^{(g,p)}(\omega,\theta_i)=\frac{\cos\theta_i-
\sqrt{\varepsilon(\omega)-\sin^2\theta_i}-
\tilde{\Pi}(\omega,\theta_i)}{\cos\theta_i+
\sqrt{\varepsilon(\omega)-\sin^2\theta_i}+
\tilde{\Pi}(\omega,\theta_i)}.
\nonumber
\end{eqnarray}
\noindent
Here, the quantities $\tilde{\Pi}_{00}$ and $\tilde{\Pi}$
are expressed via the polarization tensor of graphene in the
following way:
\begin{eqnarray}
&&
\tilde{\Pi}_{00}(\omega,\theta_i)=-\frac{ic}{\hbar\omega}\,
\frac{\cos\theta_i}{\sin^2\theta_i}\,{\Pi}_{00}(\omega,\theta_i),
\nonumber \\
&&
\tilde{\Pi}(\omega,\theta_i)=\frac{ic^3}{\hbar\omega^3}\,
\frac{1}{\sin^2\theta_i}\,{\Pi}(\omega,\theta_i),
\label{eq2}
\end{eqnarray}
\noindent
where the quantity $\Pi$ is defined as
\begin{equation}
{\Pi}(\omega,\theta_i)=\frac{\omega^2}{c^2}[\sin^2\theta_i
{\Pi}_{\rm tr}(\omega,\theta_i)+
\cos^2\theta_i{\Pi}_{00}(\omega,\theta_i)]
\label{eq3}
\end{equation}
\noindent
and ${\Pi}_{\rm tr}=\Pi_k^{\,k}$ is the trace of the polarization
tensor.

The explicit exact formulas for the quantities ${\Pi}_{\rm 00}$,
${\Pi}_{\rm tr}$ and ${\Pi}$ at real frequencies in the case of
gapped graphene are presented in Ref.~\cite{31}. Here, we use only
the approximate asymptotic representations for these formulas,
which are quite sufficient to calculate the impact of nonzero
mass-gap parameter on the reflectivity of graphene-coated plate.
In particular, we omit the nonlocal contributions which are of the
order of $v_F^2/c^2\sim 10^{-5}$.

We consider first the frequency region $\hbar\omega<\Delta$,
where quantities ${\Pi}_{\rm 00}$ and ${\Pi}$ are real. Using
Eq.~(\ref{eq2}) and the results for ${\Pi}_{\rm 00}$ and ${\Pi}$
obtained in Eqs.~(5), (6), (30), and (56) of Ref.~\cite{31}
one finds
\begin{eqnarray}
&&
\tilde{\Pi}_{00}(\omega,\theta_i)\approx 2i\alpha
\cos\theta_i\,\left[\Phi_1\left(\frac{\hbar\omega}{\Delta}\right)
+\frac{4}{\nu}I(\mu,\nu)\right],
\nonumber\\
&&\label{eq4} \\[-3mm]
&&
\tilde{\Pi}(\omega,\theta_i)\approx\tilde{\Pi}(\omega)
\approx 2i\alpha
\,\left[\Phi_1\left(\frac{\hbar\omega}{\Delta}\right)
+\frac{4}{\nu}I(\mu,\nu)\right],
\nonumber
\end{eqnarray}
\noindent
where
\begin{eqnarray}
&&
\Phi_1(x)=\frac{1}{x}-\left(1+\frac{1}{x^2}\right)
{\rm arctanh}\,{x},
\label{eq5} \\
&&
I(\mu.\nu)=\int_{\mu}^{\infty}\frac{dv}{e^v+1}\left(1+
\frac{4\mu^2+\nu^2}{4v^2-\nu^2}\right)
\nonumber
\end{eqnarray}
\noindent
and the following parameters are introduced:
\begin{equation}
\mu\equiv\frac{mc^2}{k_BT}=\frac{\Delta}{2k_BT},
\qquad
\nu\equiv\frac{\hbar\omega}{k_BT}.
\label{eq6}
\end{equation}

Note that, strictly speaking, Eq.~(56) of Ref.~\cite{31} and, thus,
Eq.~(\ref{eq4}) here, are defined under a more stringent condition
$\hbar\omega\lesssim mc^2$. Computations show, however, that
Eq.~(\ref{eq4}) actually leads to very small errors, as compared to
the exact formulas, in a wider frequency region
$\hbar\omega<2mc^2=\Delta$.
Note also that both the exact formulas and the approximate
expressions for $\tilde{\Pi}_{00}$ and $\tilde{\Pi}$ defined in
Eq.~(\ref{eq4}) are characterized by two peculiarities.
In the limiting case $\hbar\omega\to\Delta$ it holds
$|\tilde{\Pi}_{00}|,\,|\tilde{\Pi}|\to\infty$ due to
$|\Phi_1|\to\infty$. As a result, the absolute values of both
reflection coefficients turn into unity over a nonphysically
narrow frequency interval (see Ref.~\cite{31} for details).
Furthermore, for some value of frequency $\hbar\omega_0<\Delta$
it holds $\tilde{\Pi}_{00}=\tilde{\Pi}=0$. According to
Eq.~(\ref{eq1}), at $\omega=\omega_0$ graphene coating does not influence
the reflectivity properties of a dielectric plate.
At temperatures satisfying the condition $k_BT\lesssim mc^2$,
the frequency $\omega_0$ differs noticeably from $\Delta/\hbar$.
This is, however, a frequency region, where the graphene coating has
only a minor influence on the reflectivity properties of a plate
(see Sec.~III). Under a condition $mc^2<k_BT$ the value of $\omega_0$
belongs to a nonphysically narrow vicinity of the frequency
$\hbar\omega=\Delta$. For very narrow gaps ($mc^2\ll k_BT$), this
case is further discussed in Sec.~III.

Below, Eqs.~(\ref{eq1}) and (\ref{eq4})--(\ref{eq6}) are used to
calculate the TM and TE reflectivities of dielectric plates coated
with gapped graphene
\begin{eqnarray}
&&
{\cal R}_{\rm TM}(\omega,\theta_i)=
|{R}_{\rm TM}(\omega,\theta_i)|^2,
\nonumber \\
&&
{\cal R}_{\rm TE}(\omega,\theta_i)=
|{R}_{\rm TE}(\omega,\theta_i)|^2
\label{eq7}
\end{eqnarray}
\noindent
in the frequency region $\hbar\omega<\Delta$. The computational results
show that a pronounced impact of graphene coating on the reflectivity
of dielectric plates occurs only under the condition
$\hbar\omega\ll k_BT$ (see Ref.~\cite{39} for the case of a gapless graphene
coating and
Sec.~III). Taking into account that at room temperature
$k_BT\approx 0.026\,$eV, the frequency region $\hbar\omega\ll k_BT$
belongs to the area of application of Eq.~(\ref{eq4}),
$\hbar\omega<\Delta$, with exception of only the case of graphene with
very narrow gaps.

To study the influence of graphene with very small mass-gap parameter
and the limiting transition to the case of gapless graphene,
 the analytic expressions for
$\tilde{\Pi}_{00}$ and $\tilde{\Pi}$ in the frequency region
$\hbar\omega\geq\Delta$ are also required in addition to Eq.~(\ref{eq4}).
In accordance with the above
remark, it is sufficient to present them under the condition
$\hbar\omega\ll k_BT$. We, thus, have
$\Delta\leqslant\hbar\omega\ll k_BT$ or, using the notation (\ref{eq6}),
$\mu\ll 1$ and $\nu\ll 1$.

Under the conditions $\mu,\,\nu\ll 1$ and  $\hbar\omega\geqslant\Delta$
the quantities $\Pi_{00}$ and $\Pi$ are obtained in Eqs.~(5)--(7) and
(79) of Ref.~\cite{31} (note that the imaginary parts of
$\Pi_{00}$ and $\Pi$ at $T=0$ cancel with the leading contributions to
the imaginary parts of thermal corrections to these quantities).
As a result, using Eq.~(\ref{eq2}), one arrives at
\begin{eqnarray}
&&
\tilde{\Pi}_{00}(\omega,\theta_i)\approx 2i\alpha
\cos\theta_i\,\left[\Phi_2\left(\frac{\Delta}{\hbar\omega}\right)
+\frac{4}{\nu}\ln(1+e^{-\nu/2})\right],
\nonumber\\
&&\label{eq8} \\[-3mm]
&&
\tilde{\Pi}(\omega)
\approx 2i\alpha
\,\left[\Phi_2\left(\frac{\Delta}{\hbar\omega}\right)
+\frac{4}{\nu}\ln(1+e^{-\nu/2})\right],
\nonumber
\end{eqnarray}
\noindent
where
\begin{equation}
\Phi_2(x)={x}-\left(1+{x^2}\right)
{\rm arctanh}\,{x}.
\label{eq9}
\end{equation}
\noindent
Using Eqs.~(\ref{eq1}) and (\ref{eq7})--(\ref{eq9}), one can
calculate the TM and TE reflectivities of a plate coated with
gapped graphene in the frequency region
$\Delta\leqslant\hbar\omega\ll k_BT$.

At not-too-high temperatures, as compared to room temperature, in the
frequency region $\hbar\omega\ll k_BT$ the dielectric permittivities
of dielectric materials are real and approximately equal to their
static values $\varepsilon(\omega)\approx\varepsilon_0$.
Taking into account that the quantities
$\tilde{\Pi}_{00}$ and $\tilde{\Pi}$ in Eqs.~(\ref{eq4}) and (\ref{eq8})
are pure imaginary and using Eq.~(\ref{eq1}), the reflectivities (\ref{eq7})
can be expressed in the form
\begin{eqnarray}
&&
{\cal R}_{\rm TM}(\omega,\theta_i)=\frac{(\varepsilon_0\cos\theta_i -
\sqrt{\varepsilon_0-\sin^2\theta_i})^2+(\varepsilon_0-\sin^2\theta_i)
|\tilde{\Pi}_{00}(\omega,\theta_i)|^2}{(\varepsilon_0\cos\theta_i +
\sqrt{\varepsilon_0-\sin^2\theta_i})^2+(\varepsilon_0-\sin^2\theta_i)
|\tilde{\Pi}_{00}(\omega,\theta_i)|^2},
\nonumber \\
&&\label{eq10}\\[-2mm]
&&
{\cal R}_{\rm TE}(\omega,\theta_i)=\frac{(\cos\theta_i -
\sqrt{\varepsilon_0-\sin^2\theta_i})^2+
|\tilde{\Pi}(\omega,\theta_i)|^2}{(\cos\theta_i +
\sqrt{\varepsilon_0-\sin^2\theta_i})^2+
|\tilde{\Pi}(\omega,\theta_i)|^2}.
\nonumber
\end{eqnarray}
\noindent
At the normal incidence ($\theta_i=0$) the TM and TE reflectivities
coincide, as it should be.

Below these equations are used to examine the optical properties
of dielectric plates coated with gapped graphene.

\section{Influence of graphene gap on the reflectivity at normal
incidence}

In this section, we calculate the reflectivity properties of fused silica
(SiO${}_2$) plates coated with gapped graphene. As mentioned in Sec.~I,
SiO${}_2$ plates are often used as substrates for coating with graphene
\cite{40,41,42,43}. The static dielectric permittivity of SiO${}_2$ is
equal to $\varepsilon_0\approx 3.82$. For SiO${}_2$ the static value
remains applicable for all frequencies $\hbar\omega<10\,$meV \cite{45}.
At larger frequencies the optical data for $\varepsilon(\omega)$ should
be used \cite{45}.

Numerical computations for the TM and TE reflectivities of SiO${}_2$ plate
coated with gapped graphene have been performed by Eqs.~(\ref{eq4}) and
(\ref{eq10}) in the frequency region from $10^{-3}\,$meV to 10\,meV.
In this frequency region the same results were obtained using
Eqs.~(\ref{eq1}), (\ref{eq4}), and (\ref{eq7}) in combination with the
optical data for $\varepsilon(\omega)$, where they are available (i.e.,
at $\hbar\omega\geqslant 2.48\,$meV \cite{45}). At higher frequencies
($\hbar\omega>10\,$meV) Eqs.~(\ref{eq1}), (\ref{eq4}), and (\ref{eq7})
should be used in computations up to $\hbar\omega=\Delta$.
At these frequencies, however, there is no influence of either the gap
parameter or the graphene coating by itself on the reflectivity properties
of a plate.

\begin{figure}[b]
\vspace*{-8cm}
\centerline{\hspace*{2.5cm}
\includegraphics{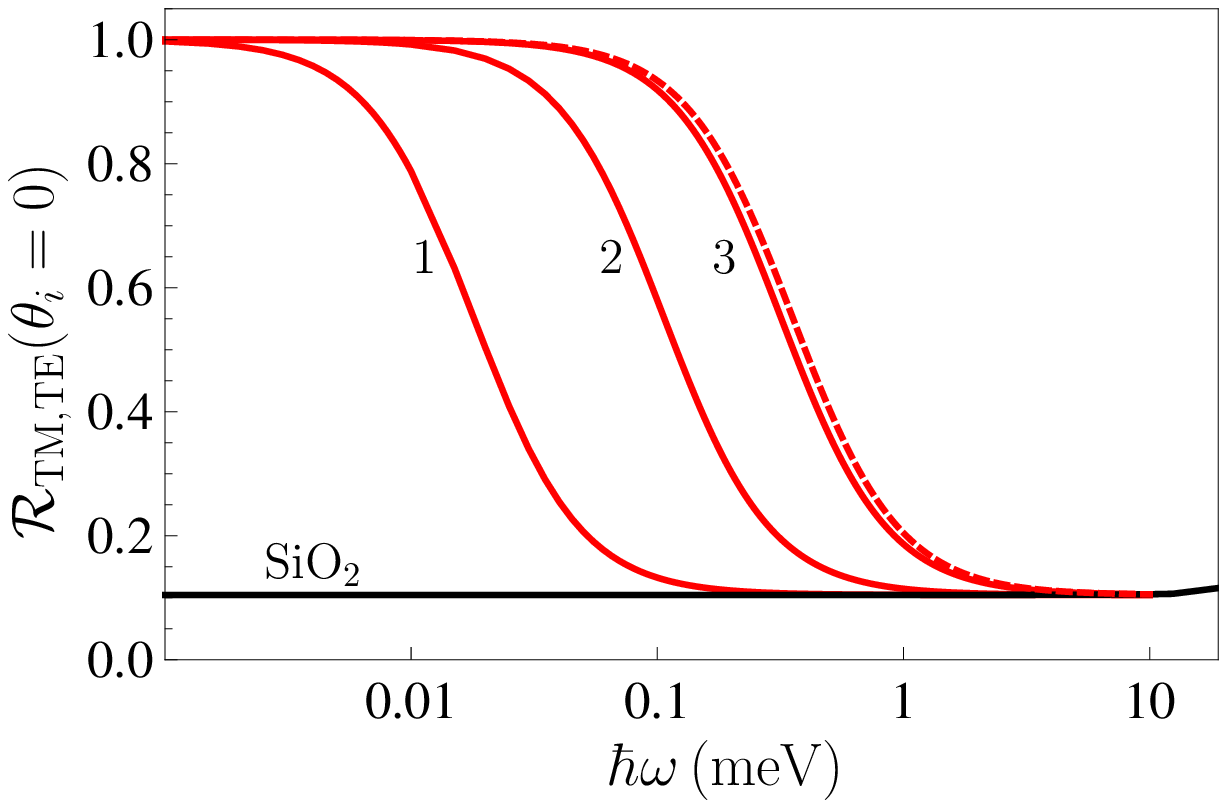}
}
\vspace*{-9cm}
\caption{\label{fg1}(Color online)
The reflectivities of graphene-coated SiO${}_2$
plates at the normal incidence are shown as functions of
frequency at $T$=300\,K by the solid lines 1, 2, and 3 for
the mass-gap parameter equal to 0.2, 0.1, and 0.02\,eV,
respectively. The solid line marked SiO${}_2$ and the
dashed line correspond to the cases of an uncoated and
coated with a gapless graphene SiO${}_2$ plate.
}
\end{figure}

In Fig.~\ref{fg1}, the reflectivities of graphene-coated SiO${}_2$ plates
at the normal incidence are shown as functions of frequency at $T=300\,$K
by the solid lines 1, 2, and 3 for the mass-gap parameter equal to
$\Delta=0.2$, 0.1, and 0.02\,eV, respectively. In the same figure, the
dashed line shows the reflectivity of a SiO${}_2$ plate
 coated with a gapless graphene at the normal
incidence.
The lower solid line marked SiO${}_2$ demonstrates the reflectivity of
an uncoated plate, which is obtained from Eq.~(\ref{eq10}) at $\theta_i=0$,
$\tilde{\Pi}_{00}=\tilde{\Pi}=0$
\begin{equation}
{\cal R}_{\rm TM,TE}=\left(
\frac{\sqrt{\varepsilon_0}-1}{\sqrt{\varepsilon_0}+1}\right)^2.
\label{eq11}
\end{equation}
\noindent
In the region $\hbar\omega>10\,$meV the frequency dependence  of
Re$\,\varepsilon(\omega)$ and nonzero Im$\,\varepsilon(\omega)$ are
of some impact on the lower solid line.

As is seen in Fig.~\ref{fg1}, the presence of a gap results in a significant decrease
of the reflectivity of graphene-coated SiO${}_2$ plate. This decrease is deeper for the
larger mass-gap parameter. As an example, for $\Delta=0.2\,$eV (which is a realistic
width of the gap for graphene-coated substrates) the reflectivity of the plate
coated with gapped graphene is smaller than that for a gapless graphene by up to
a factor of 8 depending on the value of frequency. Comparing Fig.~\ref{fg1} with
Fig.~3 of Ref.~\cite{39}, where similar results for a gapless graphene are presented
at different temperatures, one can conclude that increasing of the mass-gap parameter
at fixed temperature produces qualitatively the same effect on the reflectivity as
does decreasing temperature in the case of a coating with pure graphene.

{}From Fig.~\ref{fg1} one can see that the graphene coating ceases to influence the
reflectivity of SiO${}_2$ plate at $\hbar\omega<10\,\mbox{meV}<\Delta$ for all three
values of the width gap considered. At the same time, the values of frequency, where the
influence of graphene coating is noticeable, are much smaller than the thermal
frequency
\begin{equation}
\omega_T\equiv\frac{k_BT}{\hbar}=26\,\mbox{meV}=
3.9\times 10^{13}\,\mbox{rad/s}
\label{eq12}
\end{equation}
\noindent
at room temperature. This is in agreement with similar result for a coating made
of pure graphene \cite{39}.

Note that for not too narrow gaps considered above it holds $\mu\sim 1$ or
$\mu>1$, whereas the inequality $\hbar\omega\ll k_BT$ means that $\nu\ll 1$.
Then it follows also that $\hbar\omega\ll\Delta$. Expanding $\Phi_1(\hbar\omega/\Delta)$
defined in Eq.~(\ref{eq5}) in powers of the small parameter $\hbar\omega/\Delta$,
one obtains
\begin{equation}
\Phi\left(\frac{\hbar\omega}{\Delta}\right) \approx -\frac{4}{3}\,
\frac{\hbar\omega}{\Delta}.
\label{eq13}
\end{equation}
\noindent
If it also assumed that $\mu$ is sufficiently large (for instance, for $\Delta=0.2\,$eV
at $T=300\,$K we have $\mu\approx 3.85$), we can neglect by unity, as compared to
$e^{v}$ in the definition of $I$  in Eq.~(\ref{eq5}). Neglecting also by the small
quantity $\nu^2$ in comparison with $4\mu^2$ and $4v^2$, and calculating the integral
in Eq.~(\ref{eq5}), one arrives at \cite{31}
\begin{equation}
I(\mu,\nu)\approx 2e^{-\mu}.
\label{eq14}
\end{equation}
\noindent
Computations show that for graphene coating with $\Delta=0.1\,$eV the use
of Eq.~(\ref{eq4}) with the functions (\ref{eq13}) and (\ref{eq14}), rather than
 (\ref{eq5}), leads to the same results plotted as line 1 in Fig.~\ref{fg1}.
For smaller $\Delta$ the more exact results are obtained by using
Eqs.~(\ref{eq4}) and (\ref{eq5}).

Note that two peculiar frequencies $\hbar\omega=\Delta$, where the reflectivity
turns into unity, and $\hbar\omega_0<\Delta$, where the graphene coating does not
influence the reflectivity properties of a plate (see Sec.~II) are not shown in
Fig.~\ref{fg1} because they occur at higher frequencies (at 20\,meV and at slightly
smaller frequency, respectively, for the line 3 related to the case of the smallest
mass-gap parameter). All these frequencies belong to the region of absorption bands
of SiO${}_2$, where the graphene coating does not make any impact on the reflectivity
properties taking into account nonphysically small half-widths of the corresponding
resonances \cite{31}.

\begin{figure}[b]
\vspace*{-8cm}
\centerline{\hspace*{2.5cm}
\includegraphics{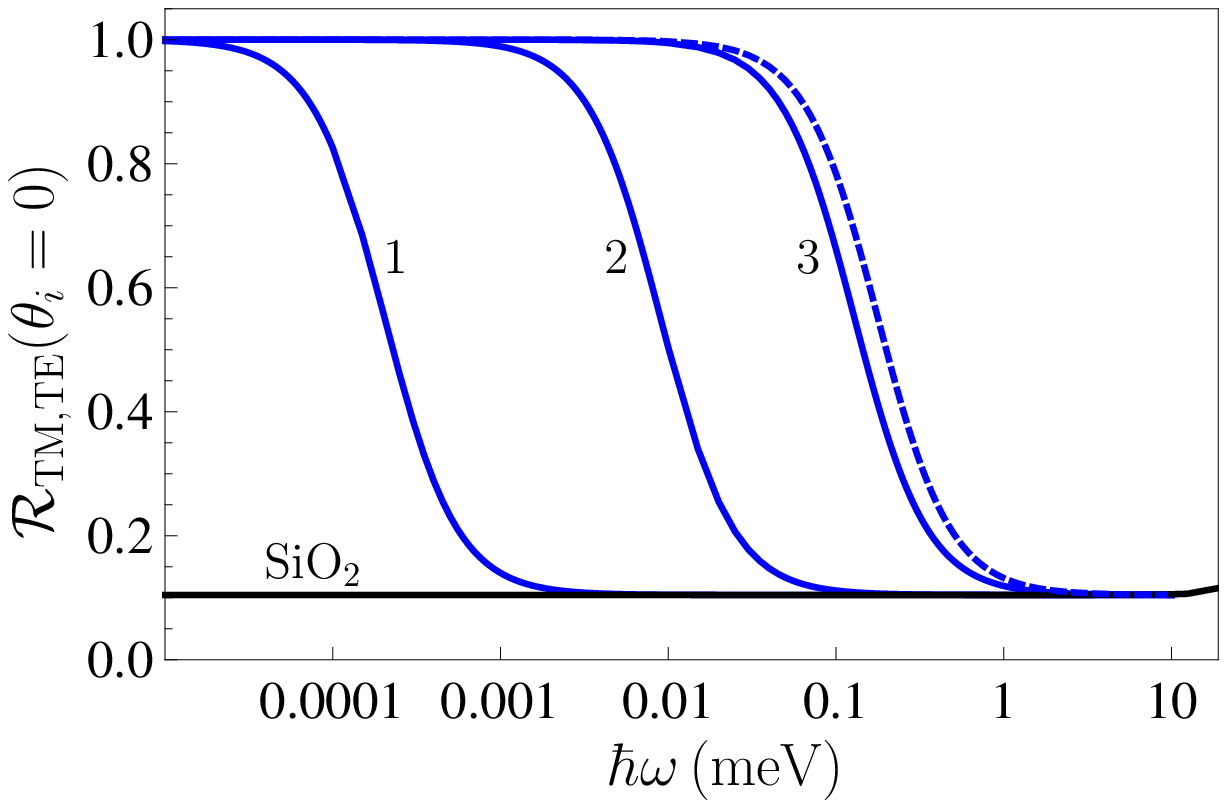}
}
\vspace*{-9cm}
\caption{\label{fg2}(Color online)
The reflectivities of graphene-coated SiO${}_2$
plates at the normal incidence are shown as functions of
frequency at $T$=150\,K by the solid lines 1, 2, and 3 for
the mass-gap parameter equal to 0.2, 0.1, and 0.02\,eV,
respectively. The solid line marked SiO${}_2$ and the
dashed line correspond to the cases of an uncoated and
coated with a gapless graphene SiO${}_2$ plate.
}
\end{figure}

To discuss the role of different values of temperature, in Fig.~\ref{fg2} we plot
the reflectivities of graphene-coated SiO${}_2$ plates at $T=150\,$K.
The same notations for different lines, as in Fig.~\ref{fg1}, are used.
In the frequency region from $10^{-5}$ to 10\,meV computations have been performed
using Eqs.~(\ref{eq4}) and (\ref{eq10}). As is seen in Fig.~\ref{fg2}, the impact of
graphene coating with different values of the mass-gap parameter is qualitatively
the same. as in Fig.~\ref{fg1}. One can conclude, however, that at $T=150\,$K the
decrease of reflectivities down to the reflectivity of an uncoated  SiO${}_2$ plate
starts at much lower frequencies than at $T=300\,$K. For example, for the line 1
($\Delta=0.2\,$eV) this decrease starts at $\hbar\omega\approx 10^{-5}\,$meV,
whereas in Fig.~\ref{fg1} it starts at for orders of magnitude higher frequency.
By and large, from the comparison of Fig.~\ref{fg2} and Fig.~\ref{fg1} it is seen
that at $T=150\,$K the impact of graphene coating with some fixed gap parameter
on the reflectivity of SiO${}_2$ plate in stronger than at $T=300\,$K.

\begin{figure}[b]
\vspace*{-8cm}
\centerline{\hspace*{2.5cm}
\includegraphics{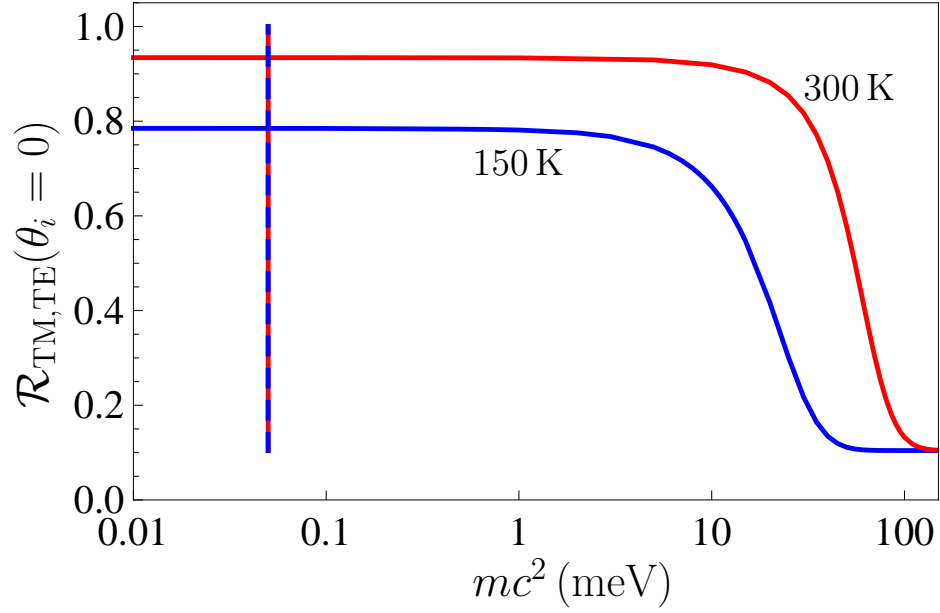}
}
\vspace*{-9cm}
\caption{\label{fg3}(Color online)
The reflectivities of graphene-coated SiO${}_2$
plates at the normal incidence are shown as functions of
quasiparticle mass at $\hbar\omega$=0.1\,meV by the solid
lines plotted at $T$=300\,K and $T$=150\,K. The overlapping
vertical solid and dashed lines indicate the value of
$mc^2$=0.05\,meV, where the reflectivity turns into unity
at $T$=300\,K and $T$=150\,K, respectively.
}
\end{figure}

In the end of this section, we calculate the reflectivity of a SiO${}_2$ plate
coated with gapped graphene at the normal incidence as a function of the
quasiparticle mass at some fixed frequency, $\hbar\omega=0.1\,$meV, for instance.
Computations are performed by using Eqs.~(\ref{eq4}) and (\ref{eq10}) in the
region of masses $2mc^2=\Delta>\hbar\omega$ and by Eqs.~(\ref{eq8}) and (\ref{eq10})
in the region $2mc^2\leqslant\hbar\omega$. In Fig.~\ref{fg3} the computational results
for the reflectivities are presented by the upper and lower solid lines computed
at 300 and 150\,K, respectively. The overlapping solid and dashed vertical lines
demonstrate the narrow resonances which occur at $T=300$ and 150\,K under the
condition $2mc^2=\hbar\omega=0.1\,$meV.

As is seen in Fig.~\ref{fg3}, for small masses below 0.1\,meV there is no influence
of the gap on the reflectivity of a graphene-coated SiO${}_2$ plate. At $T=150\,$K
the gap comes into play at smaller width than at $T=300\,$K. As a result,
the reflectivity of a graphene-coated plate drops to that of an uncoated graphene for
$mc^2\approx 65\,$meV at $T=150\,$K and for $mc^2\approx 140\,$meV at $T=300\,$K.

Quantitatively, from Fig.~\ref{fg3} at $T=300\,$K one has
${\cal R}_{\rm TM,TE}=0.934$ for $m=0$ and 0.919 for $m=0.01\,$meV, i.e.,
the relative change of only 1.6\%. At $T=150\,$K one obtains
${\cal R}_{\rm TM,TE}=0.784$ for $m=0$ and 0.662 for $m=0.01\,$meV. This corresponds
to a much larger relative change of 18.4\%.  Thus, at lower temperature the graphene
coating with the same gap produces a larger relative impact on the reflectivity,
as compared to the case of gapless graphene, in accordance with a conclusion
made from the comparison of Figs.~\ref{fg1} and \ref{fg2}.

\section{Influence of graphene coating on the angle dependence of reflectivities}

Here, we consider the angle dependences of both TM and TE reflectivities
of dielectric plates coated with
gapped graphene with different values of the mass-gap parameter. Computations are
performed at $T=300\,$K by using Eqs.~(\ref{eq4}) and (\ref{eq10}), i.e., in the
region, where the nonzero gap makes an important impact on the reflectivity properties
of graphene-coated plates.

\begin{figure}[b]
\vspace*{-3cm}
\centerline{\hspace*{2.5cm}
\includegraphics{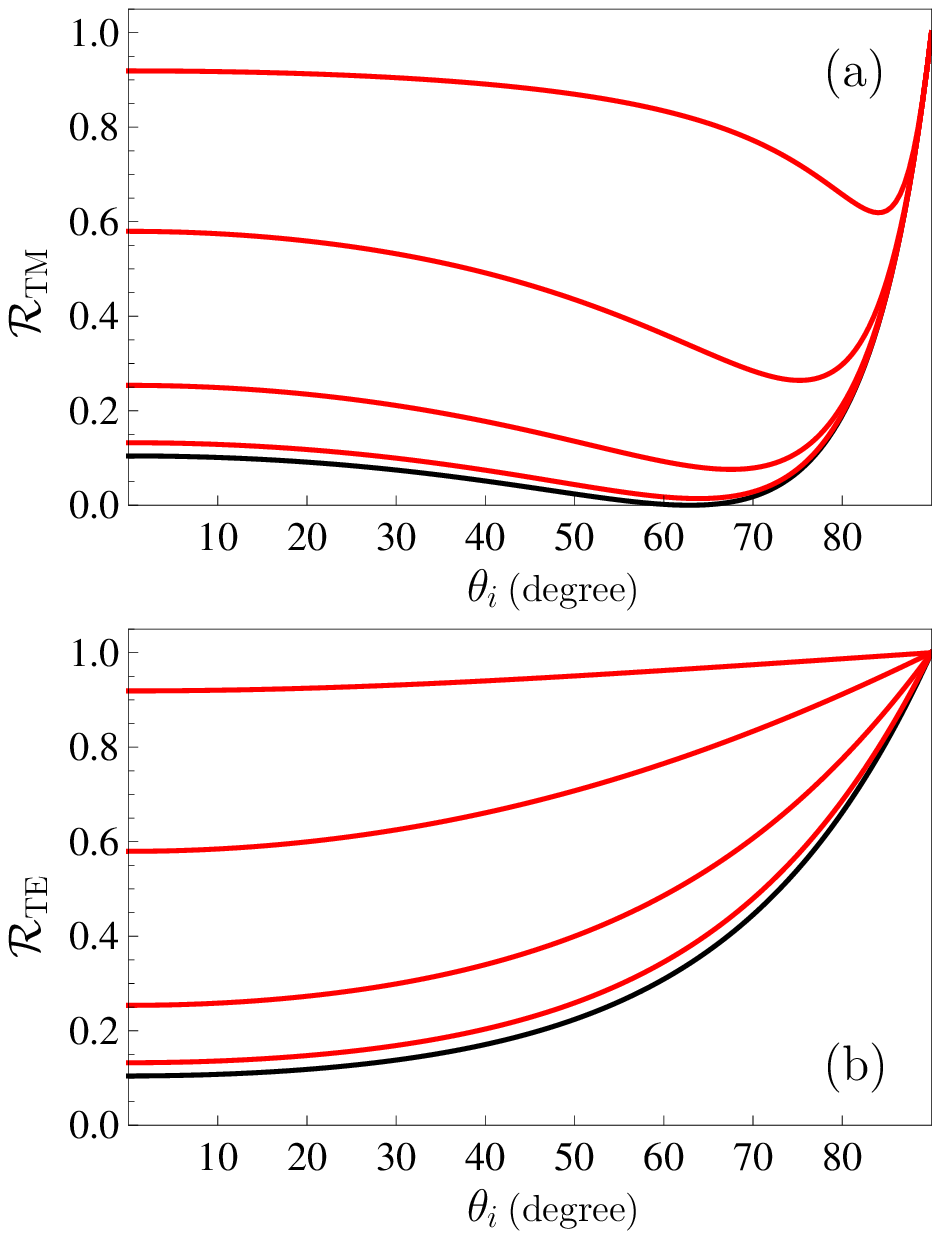}
}
\vspace*{-13cm}
\caption{\label{fg4}(Color online)
The reflectivities of SiO${}_2$ plates with no
graphene coating and coated by gapped graphene with
$\Delta$=0.2, 0.15, 0.1, and $\leqslant$0.02\,eV are shown as
functions of an incidence angle at $T$=300\,K,
$\hbar\omega$=0.1\,meV by the five solid lines from bottom
to top, respectively, for (a) TM polarization and (b) TE
polarization of the incident light.
}
\end{figure}

In Fig.~\ref{fg4}(a) the TM reflectivities of SiO${}_2$ plates with no graphene
coating and coated by graphene with $\Delta=0.2$, 0.15, 0.1, and $\leqslant 0.02\,$eV are
shown at $\hbar\omega=0.1\,$eV as functions of the incidence angle by the five solid
lines from bottom to top, respectively. In Fig.~\ref{fg4}(b) similar computational
results for the TE reflectivity are presented.

As is seen in Fig.~\ref{fg4}(a), the most pronounced influence of graphene coating on
the TM reflectivity is for a gapless or having a narrow gap
$\Delta\leqslant 0.02\,$eV graphene (the upper line).
With increasing width of the gap, the reflectivity
becomes smaller, preserving the character of an angle dependence with a typical
minimum value which shifts to smaller angles of incidence. Only for an uncoated
SiO${}_2$ plate (the lowest line) the TM reflectivity takes zero value at the Brewster
angle $\theta_B=62.90^{\circ}$. In this case the reflected light is fully (TE) polarized.
Note that at much higher frequencies the full polarization of reflected light is
possible also for a SiO${}_2$ plate coated with gapless graphene \cite{39}. In this
frequency region, however, the presence of nonzero gap does not influence the reflectivity
properties of graphene-coated plate.

{}From  Fig.~\ref{fg4}(b) it is seen that the TE reflectivities of graphene-coated
SiO${}_2$ plates also decrease with increasing $\Delta$ at any angle of incidence.
Unlike Fig.~\ref{fg4}(a), however, the TE reflectivities decrease monotonously with
decreasing angle of incidence.

\begin{figure}[b]
\vspace*{-3cm}
\centerline{\hspace*{1.5cm}
\includegraphics{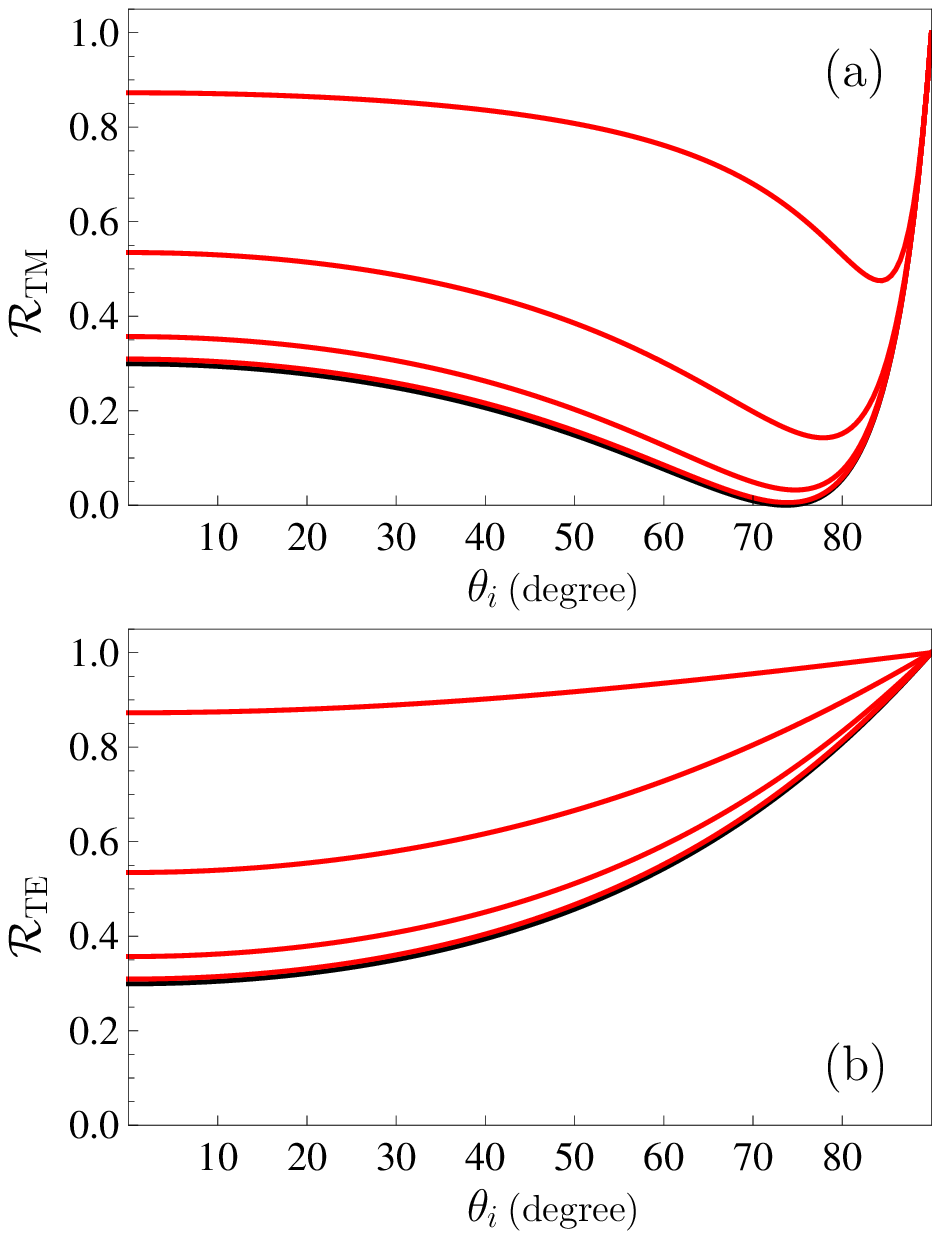}
}
\vspace*{-13cm}
\caption{\label{fg5}(Color online)
The reflectivities of Si plates with no
graphene coating and coated by gapped graphene with
$\Delta$=0.2, 0.15, 0.1, and $\leqslant$0.02\,eV are shown as
functions of an incidence angle at $T$=300\,K,
$\hbar\omega$=0.1\,meV by the five solid lines from bottom
to top, respectively, for (a) TM polarization and (b) TE
polarization of the incident light.
}
\end{figure}

To investigate the role of material properties of a substrate, similar computations of
angle dependences of the reflectivities have been performed for graphene-coated plates
made of high-resistivity Si. This material possesses much higher static dielectric
permittivity $\varepsilon_0=11.67$ \cite{45}. The computational results obtained using
Eqs.~(\ref{eq4}) and (\ref{eq10}) are presented in Fig.~\ref{fg5} using the same
notations and the same values of all parameters as in Fig.~\ref{fg4}.

As is seen in Fig.~\ref{fg5}(a,b), the dependences of both TM and TE reflectivities
of Si plates coated with gapped graphene are qualitatively the same as for graphene-coated
SiO${}_2$ plates [see Fig.~\ref{fg4}(a,b)]. Specifically, the Brewster angle
$\theta_B=73.68^{\circ}$  is achieved for only an uncoated Si plate. It is seen, however,
that  for both TM and TE reflectivities the two lowest lines (related to the cases of an
uncoated plate and of a plate coated by graphene with $\Delta=0.2\,$eV) are overlapping.
This means that coating by graphene with relatively wide gap does not make a dramatic
effect on the angle dependences of reflectivities of plates with large $\varepsilon_0$.

It is interesting to investigate a dependence of the angle $\theta_0$, where the minimum
TM reflectivity of graphene-coated plate is achieved [see Figs.~\ref{fg4}(a) and \ref{fg5}(a)],
on the quasiparticle mass $m$ for different plate materials at different temperatures.
{}From the first equality in Eq.~(\ref{eq10})  one concludes that the quantity
\begin{equation}
F(\omega)\equiv\frac{|\tilde{\Pi}_{00}(\omega,\theta_i)|}{\cos\theta_i}
\label{eq15}
\end{equation}
\noindent
does not depend on the angle of incidence. Then, from the first equality in Eq.~(\ref{eq10}),
it is easily seen that the minimum value of the quantity ${\cal R}_{\rm TM}$ is achieved at
the angle of incidence $\theta_i=\theta_0$ satisfying the following equation:
\begin{eqnarray}
&&
\cos^6\theta_0+\frac{3(\varepsilon_0-1)}{2}\cos^4\theta_0
\label{eq16}\\
&&
+
\frac{(\varepsilon_0-1)^2[\varepsilon_0+1+F^2(\omega)]}{2F^2(\omega)}\cos^2\theta_0
-\frac{(\varepsilon_0-1)^2}{2F^2(\omega)}=0.
\end{eqnarray}
\noindent
This equation has only one solution for $\cos^2\theta_0$  satisfying the condition
\begin{equation}
0\leqslant\cos^2\theta_0\leqslant1.
\label{eq17}
\end{equation}

\begin{figure}[b]
\vspace*{-8cm}
\centerline{\hspace*{2.5cm}
\includegraphics{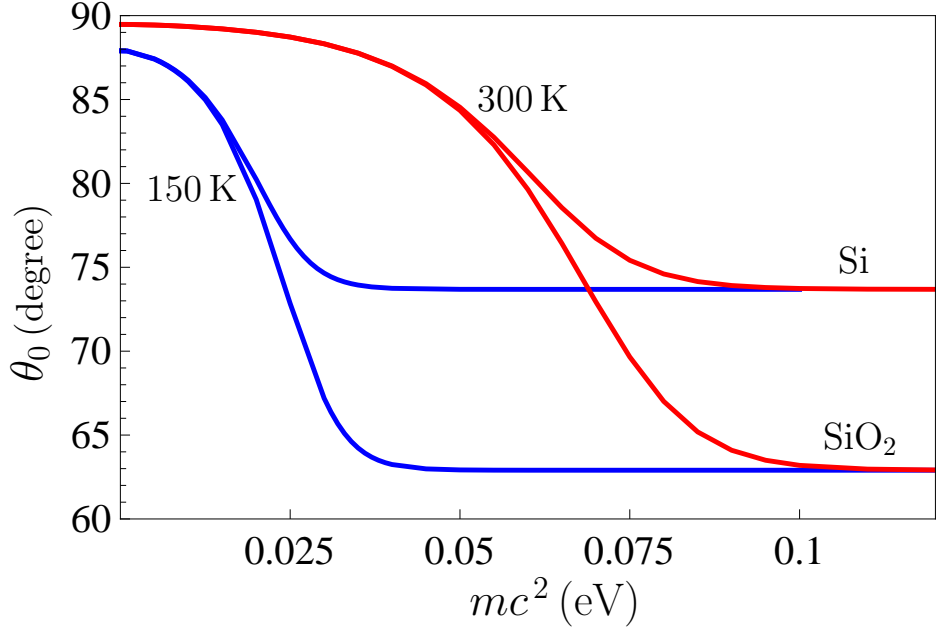}
}
\vspace*{-9cm}
\caption{\label{fg6}(Color online)
The incidence angles such that the TM
reflectivities of graphene-coated Si and SiO${}_2$ plates
at $T$=300K and $\hbar\omega$=0.1\,meV take the minimum
values, are shown as functions of quasiparticle mass by
the top pair of upper and lower lines, respectively. The
lower pair of lines presents similar results computed at
$T$=150\,K.
}
\end{figure}

In Fig.~\ref{fg6}  we present computational results for the angle $\theta_0$,
as a function of the quasiparticle mass $m$, obtained by solving Eq.~(\ref{eq16}).
The pair of two upper lines is plotted at $T=300\,$K, $\hbar\omega=0.1\,$meV
for Si and SiO${}_2$ plates from top to bottom, respectively.
The pair of two lower lines is plotted at $T=150\,$K, $\hbar\omega=0.1\,$meV
for the same materials in the same sequence. As is seen in Fig.~\ref{fg6},
at small masses (narrow gaps) the values of $\theta_0$ depend only on the
temperature and are almost independent on the plate material. After some
transition region of masses, where both the value of temperature and material
properties influence the result, the value of $\theta_0$ is coming to depend
of only on the plate material. Specifically, for SiO${}_2$ plate at $T=300\,$K
the angle $\theta_0$ is equal to $89.48^{\circ}$ for a gapless graphene coating
and takes the values $84.35^{\circ}$ and $63.19^{\circ}$ for gapped graphene
coatings with $m=0.05$ and 0.1\,eV, respectively. For graphene-coated Si plate
at the same temperature one has $\theta_0=89.48^{\circ}$,  $84.54^{\circ}$,
and $73.74^{\circ}$  $m=0$, 0.05, and 0.1\,eV, respectively.
At $T=150\,$K for the same respective values of $m$ the angle $\theta_0$ is
equal to $87.90^{\circ}$,  $62.92^{\circ}$, and $62.90^{\circ}$ for SiO${}_2$
and $87.90^{\circ}$,  $73.69^{\circ}$, and $73.68^{\circ}$ for Si plate. 
As is seen in Fig. 6, for the plate materials with larger $\varepsilon_0$, 
the value of $\theta_0$ becomes mass-independent for smaller values of the 
mass-gap parameter of graphene. One can also see that this property is 
preserved at any temperature.

\section{Conclusions and discussion}

In this paper, we have investigated the optical properties
of dielectric plates coated with gapped graphene on the
basis of first principles of quantum electrodynamics. For
this purpose, graphene was described by the polarization
tensor with a nonzero mass-gap parameter defined along
the real frequency axis at any temperature, whereas the
plate material was characterized by its dielectric
permittivity. Simple approximate expressions for the
polarization tensor and for the transverse magnetic and
transverse electric reflectivities of graphene-coated
plates have been derived. These expression are valid for
all frequencies smaller than the width of the gap, where
the mass-gap parameter makes a major impact on the
reflectivities, and also for frequencies exceeding the
width of the gap (the latter under a condition that they
are much smaller than the thermal frequency).

The obtained expressions have been applied to calculate an
impact of the mass-gap parameter on the reflectivity of
graphene-coated SiO${}_2$ plates at the normal incidence
at different temperatures. It is shown that at fixed
frequency the reflectivity of graphene-coated plates
decreases with increasing mass-gap parameter. Calculations
show that for a wider gap the reflectivity begins its
decreasing from unity to the reflectivity of an uncoated
SiO${}_2$ plate at lower frequency. For the gap width
$\Delta$=0.02 eV at $T$=300K the reflectivities of plates
coated with gapped and gapless graphene are shown to be
rather close. At the same time, the reflectivity of
SiO${}_2$ plate coated with gapped graphene with
$\Delta$=0.2 eV may be smaller by up to a factor of 8 than
the reflectivity of a plate coated with gapless graphene.
Qualitatively the dependences of reflectivities on
frequency are akin in the cases of graphene coatings with
different values of the mass-gap parameter and a gapless
graphene coating at different temperatures. The dependence
of a reflectivity on the quasiparticle mass $m$ is also
computed over the wide range of $mc^2$ from $10^{-2}$ meV
to 100 meV at different temperatures.

Furthermore, the obtained expressions have been used to
calculate the angle dependences of the TM and TE
reflectivities of SiO${}_2$ and Si plates coated with
gapped graphene with different values of the mass-gap
parameter. It was shown that the TM reflectivity has a
minimum value at some incidence angle $\theta_0$, which
depends on the mass-gap parameter of graphene coating.
These minimum values, however, are always larger than zero.
Thus, the reflected light from a plate coated with gapped
graphene cannot be fully polarized. This is different from
the coating with a gapless graphene where the full (TE)
polarization of reflected light is possible at high
frequencies. We have also calculated the dependence of an
angle $\theta_0$ on the mass-gap parameter for both
SiO${}_2$ and Si plates at different temperatures. It is
shown that $\theta_0$ decreases with increasing mass-gap
parameter and that the limiting values of $\theta_0$ at
large $m$ do not depend on the temperature, but only on
the plate material.

The above results demonstrate that a nonzero gap of
graphene has profound effects on the optical properties
of graphene-coated dielectric plates, which should be
taken into account in numerous applications, such as the
optical detectors, solar cells, optoelectronic switches
and others mentioned in Sec. I.
In Sec. I it was also discussed that a nonzero gap arises 
under the influence of different factors including the defects
of structure or impurities. A more full account of the role of impurities 
can be made by considering nonzero chemical potential. 
A rough estimate \cite{49a} shows that the above results 
obtained for zero chemical potential can be valid for the 
concentration of impurities below
 $1.2\times 10^{10}\,\mbox{cm}^{-2}$.
 In this respect further
generalization of the developed formalism, e.g., for the
case of graphene coatings with nonzero chemical potential
\cite{46} is of interest for future work.

%%%%%%%%%%%%%%%%%%%%%%%%%%%%%%%%%%%%%%%%%%%%%%%%%%%%%%%%%%
\section*{Acknowledgments}

The work of V.M.M.~was partially supported by the Russian Government
Program of Competitive Growth of Kazan Federal University.

%%%%%%%%%%%%%%%%%%%%%%%%%%%%%%%%%% %%%%%%%%%%%%%%%%%%%%%%%%%%%%%%%%%%%%%%%%%%%%%%%%%%%
%%%%%%%%%%%%%%%%%%%%%%%%%%%%

%%%%%%%%%%%%%%%%
%\end{document}
%%%%%%%%%%%%%%%
%%%%%%%%%%%%%%%
\newpage
%%%%%%%__FIGURE__1__%%%%%%%%%%%%%%%%%%%%
%%%%%%%%%%%%%
%%%%%%%__FIGURE__2__%%%%%%%%%%%%%%%%%%%%
%%%%%%%%%%%%%
%%%%%%%__FIGURE__3__%%%%%%%%%%%%%%%%%%%%
%%%%%%%__FIGURE__4__%%%%%%%%%%%%%%%%%%%%
%%%%%%%%%%%%%
%%%%%%%__FIGURE__5__%%%%%%%%%%%%%%%%%%%%
%%%%%%%%%%%%%
%%%%%%%__FIGURE__6__%%%%%%%%%%%%%%%%%%%%
%%%%%%%%%%%%%
%%%%%%%%%%%%%
\end{document}